\documentclass[prb,twocolumn,superscriptaddress,showpacs,amsmath,amssymb]{revtex4-2}
\usepackage[colorlinks=true, linkcolor=blue, filecolor=blue, urlcolor=blue, citecolor=blue]{hyperref}
\usepackage{amsfonts}
\usepackage{bbm}
\usepackage{graphicx}
\usepackage{dcolumn}
\usepackage{bm}
\usepackage{ulem}

\begin{document}
	
	\title{Design of Josephson Diode Based on Double Magnetic Impurities}
	
	\author{Yu-Fei Sun}
	\affiliation{International Center for Quantum Materials, School of Physics, Peking University, Beijing 100871, China}
	\affiliation{Hefei National Laboratory, Hefei 230088, China}
	
	\author{Yue Mao}
	\email[]{maoyue@stu.pku.edu.cn}
	\affiliation{International Center for Quantum Materials, School of Physics, Peking University, Beijing 100871, China}
	\affiliation{Hefei National Laboratory, Hefei 230088, China}
	
	\author{Qing-Feng Sun}
	\email[]{sunqf@pku.edu.cn}
	\affiliation{International Center for Quantum Materials, School of Physics, Peking University, Beijing 100871, China}
	\affiliation{Hefei National Laboratory, Hefei 230088, China}
	\affiliation{Collaborative Innovation Center of Quantum Matter, Beijing 100871, China}

\begin{abstract}
We theoretically propose a universal superconducting diode device based on double magnetic impurities which are coupled to the connection region of the Josephson junction.
The positive and negative currents flowing across the junction can generate opposite magnetic fields, flipping the magnetic moment of the side magnetic impurity to the opposite directions, and in turn, the two impurities will have different impacts on the opposite currents.
This results in the phenomenon that the positive and negative critical currents are unequal, referred to as the superconducting diode effect (SDE). Using the nonequilibrium Green's function method (NEGF), we obtain the direction-dependent critical currents. We confirm the emergency of the SDE and demonstrate the dependence of the superconducting diode efficiency on a range of parameters including the magnitude, the direction, and the position of the magnetic moment. Besides, we systematically analyze the symmetry relations of the nonreciprocity in our system. Our proposal has high practicability by avoiding the demand on the external magnetic field, the Cooper pair momentum, and the spin-orbit coupling. Our approach opens up new possibilities for the development of nonreciprocal electronic circuits and provides an alternate perspective on the advancement of superconducting devices.
\end{abstract}
\maketitle
	
\section{\label{sec1}Introduction}

The traditional semiconductor diode, which has high resistance in one direction and low resistance in the opposite direction,
has played a crucial role in modern electronic devices \cite{sze_semiconductor_1985,fruchart_nonreciprocal_2021,shockley_theory_1949}. The applications of its non-reciprocal transport include current rectifiers, voltage-controlled oscillators, solar cells, and so on.
However, since there are inevitably nonzero resistances in the diodes,
the issue of energy dissipation in modern integrated circuits remains an urgent challenge that requires resolution.
In analogy to the semiconductor diode, the superconducting diode effect (SDE) allows current to flow as supercurrent in one direction with zero resistance, but as normal current in the opposite direction. Utilizing the supercurrent, the SDE can help to achieve more energy-efficient and dissipationless devices. It can play a significant role in the fields of the basic component of computer memory and the future quantum devices such as quantum sensors and logic circuits \cite{tinkham_introduction_2004,braginski_superconductor_2019,linder_superconducting_2015,liu_2d_2019}.
	
To realize the SDE, various proposals have been explored experimentally and theoretically \cite{banerjee_altermagnetic_2024,chen_intrinsic_2024,daido_superconducting_2022,depicoli_superconducting_2023,PhysRevB.110.014518,hou_ubiquitous_2023,kealhofer_anomalous_2023,nakamura_orbital_2024,roig_superconducting_2024,zhai_prediction_2022,zhang_fabryperot_2024,zinkl_symmetry_2022,ando_observation_2020,nagaosa_nonreciprocal_2024,tokura_nonreciprocal_2018,wakatsuki_nonreciprocal_2018,daido_intrinsic_2022,davydova_universal_2022,he_phenomenological_2022,ilic_theory_2022,legg_superconducting_2022,yuan_supercurrent_2022,costa_microscopic_2023,hu_josephson_2023,kokkeler_fieldfree_2022,osin_anomalous_2024,pillet_josephson_2023,sinner_superconducting_2024,volkov_josephson_2024,wang_efficient_2024,cheng_josephson_2023,debnath_gatetunable_2024,sun_design_2023,lu_tunable_2023,soori_nonequilibrium_2023,cayao_enhancing_2024,legg_parityprotected_2023,baumgartner_effect_2022,lotfizadeh_superconducting_2024,pal_josephson_2022,lin_zerofield_2022,wu_fieldfree_2022,trahms_diode_2023,qi_hightemperature_2025}. The initial experimental observation of SDE was achieved by applying an external magnetic field in an artificial superlattice [Nb/V/Ta] \cite{ando_observation_2020}.
Typically, it is required to break both the inversion and time-reversal symmetry of the system to achieve SDE \cite{nagaosa_nonreciprocal_2024,tokura_nonreciprocal_2018,wakatsuki_nonreciprocal_2018}. The breaking of inversion symmetry is usually achieved by spin-orbit interactions, while the breaking of time-reversal symmetry is usually achieved by applying a magnetic field. The combination of spin-orbit interaction and magnetic field can cause a finite Cooper pair momentum that results in the SDE \cite{daido_intrinsic_2022,davydova_universal_2022,he_phenomenological_2022,ilic_theory_2022,legg_superconducting_2022,yuan_supercurrent_2022}. In addition, Josephson junction is an alternative platform to achieve the SDE, also called Josephson diode effect (JDE) \cite{costa_microscopic_2023,hu_josephson_2023,kokkeler_fieldfree_2022,osin_anomalous_2024,pillet_josephson_2023,sinner_superconducting_2024,volkov_josephson_2024,wang_efficient_2024,cheng_josephson_2023,debnath_gatetunable_2024,sun_design_2023,lu_tunable_2023,soori_nonequilibrium_2023,cayao_enhancing_2024,legg_parityprotected_2023,baumgartner_effect_2022,lotfizadeh_superconducting_2024,pal_josephson_2022,lin_zerofield_2022,wu_fieldfree_2022,trahms_diode_2023}.
So far, JDE has been proposed or realized in various Josephson junctions, which usually has demand for strong spin-orbit interaction \cite{lotfizadeh_superconducting_2024,pal_josephson_2022}.
Theoretically, this non-reciprocal behavior can be seen by the current-phase relationship (CPR) with asymmetry between the critical currents in opposite directions \cite{cheng_josephson_2023,sun_design_2023}.
To better enhance the application of superconducting diodes in future quantum devices, reaching field-free JDEs and reducing special requirements for the materials and designs are essential \cite{lin_zerofield_2022}.
In 2022, Wu et al. \cite{wu_fieldfree_2022} successfully achieved the magnetic-field-free SDE in Josephson junction, which attracted lots of attention and stimulated further research without applying a magnetic field.
	
Very recently, Trahms et al. \cite{trahms_diode_2023} experimentally achieved that JDE can be induced by a very simple structure: only inserting a magnetic atom into a Josephson junction. It opens a new avenue for creating superconducting diodes with low demand.
They observe that the incorporation of a single magnetic atom can lead to JDE in the junction, with remarkable asymmetry in the critical current.
At the same period, we theoretically proposed JDE by coupling magnetic impurity to the connection region of Josephson junction \cite{sun_design_2023}, which also incidentally gives an explanation on the above experimental work.
We proposed a mechanism: the Josephson currents through the junction generate magnetic fields, which can have influence on the magnetic moment amplitude of magnetic impurity.
For currents in positive and negative directions, the induced magnetic fields are opposite, with different influences on the magnetic moment.
Thus, the critical currents in opposite directions are unequal.
Our previous work put forward a new perspective to explain the origin of JDE.
In that work, the influence of magnetic field on the magnetic moment is regulating the magnitude of the magnetic moment, corresponding to a hard magnet \cite{krishnan_fundamentals_2016}.
However, in soft magnets, a small magnetic field can immediately flip the \textit{direction} of magnetic moment to parallel to the magnetic field, causing a significant change of the direction of the magnetic moment \cite{krishnan_fundamentals_2016}.
In comparison, the change in the magnitude of the magnetic moment is relatively insignificant.
Is it feasible to achieve JDE with high efficiency and flexible controllability, based on the sharp flip of magnetic moment while its magnitude remains unchanged?

In this paper, we theoretically propose a Josephson diode based on double magnetic impurities, utilizing the flip of magnetic moment: two magnetic impurities exist in the superconductor-normal conductor-superconductor (SNS) junction, as shown in Fig. \ref{Fig1}.
One is in the center of the normal conductor, and the other one is in the side.
When a Josephson current flows through the junction, a magnetic field is generated. For the center magnetic impurity, the currents flowing on its two sides are symmetric, and the magnetic fields generated by the currents on the two sides are canceled out.
In fact, as early as 30 years ago, the manipulation at atomic precision has been achieved \cite{crommie_confinement_1993}, thus it is feasible to precisely put the magnetic impurity in the center.
Even with a deviation, it is slight. 
It will not affect the result that the magnetic fields on the two sides can still be almost canceled out.
Thus, the center magnetic moment feels almost zero magnetic field and keeps its direction unchanged by the current direction.
On the contrary, for the side magnetic impurity, the net magnetic fields generated by the currents is nonzero.
Therefore, the magnetic moment of the side magnetic impurity is flipped by the current-induced magnetic field and points to opposite directions for opposite currents [see Fig. \ref{Fig1}(a,b)]. The Josephson currents in opposite directions have different influences on the side magnetic moment, and in turn, different magnetic moments will have different influences on the currents.
So, the critical currents in opposite directions are unequal, and the JDE emerges.
The main innovation compared to our previous paper \cite{sun_design_2023} is the different type of magnetic material considered. 
In the previous study, the magnetic impurity was based on hard magnets, where the magnetic moment is difficult to flip and its magnitude changes with the magnetic field. In contrast, this paper focuses on soft magnets, where the magnetic moment can be easily flipped, and its magnitude is constant \cite{krishnan_fundamentals_2016}.
In this case, only one magnetic impurity with the constant magnitude can not induce JDE, because the direction of single magnetic moment has no influence on the current.
Using the nonequilibrium Green's function methods, we calculate the Josephson current of the system and obtain CPRs to confirm the emergence of JDE.
We thoroughly study how the current and the superconducting diode efficiency rely on system parameters. In addition, we perform the universal symmetry analysis to reflect the behavior of the current.
The results demonstrate that JDE does emerge in our system and reaches a significant efficiency that can be conveniently adjusted by the properties of the magnetic moment.
Besides, our proposal has very low requirements: the finite-momentum Cooper pairs in unconventional superconductors, the spin-orbit coupling, and the external magnetic field are all not required. 
The only requirement is that two magnetic impurities 
are placed in the center and side respectively during the construction of the device.
Furthermore, our work provides a theoretically concise mechanism to achieve JDE, which is fundamental and practicable for future applications.

The rest of this paper is organized as follows. In Sec. \ref{sec2}, we present the Hamiltonians for our Josephson diode device and calculate the Josephson current by the nonequilibrium Green's function method in the lattice model.
In Sec. \ref{sec3}, the numerical results are calculated for the Josephson current, and the origins of JDE are explained.
In Sec. \ref{sec4}, we thoroughly study the influence of the properties of the magnetic moments on JDE. Symmetry relations are also systematically derived. Finally, discussion and conclusion are given in Sec. \ref{sec5}. 
The discretization of the Hamiltonian is given in the Appendix.

\section{\label{sec2}Model and formulations}
	
\begin{figure}[!htb]
    \centerline{\includegraphics[width=\columnwidth]{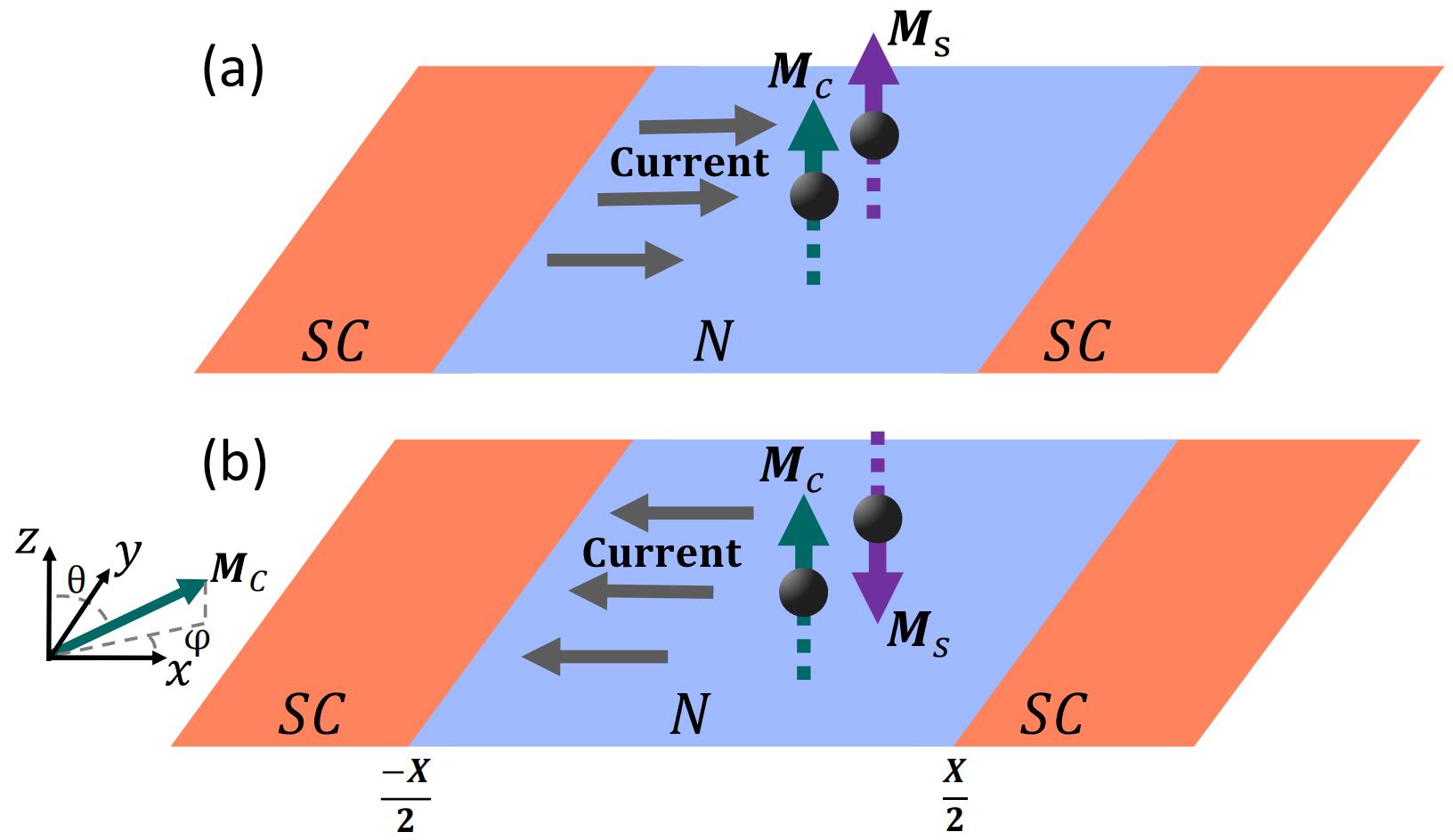}}
    \caption{
Schematic illustration of the Josephson diode.
	In the two-dimensional SNS Josephson junction,
a magnetic impurity is set at the center of the normal conductor (shown as the green vector $\mathbf{M}_c$).
The other magnetic impurity is set at the side of the normal conductor (shown as the purple vector $\mathbf{M}_s$). The direction of $\mathbf{M}_s$ is flipped by the current-induced magnetic field, and is determined by the direction of the current (shown as the gray vector).
	For a positive current, $\mathbf{M}_s$ is at $+z$ direction as shown in (a), while for a negative current $\mathbf{M}_s$ is at $-z$ direction as shown in (b).
The opposite currents have different effects on the magnetic moments,
which in turn leads to different critical currents in opposite directions, i.e. the JDE. The magnetic moment of center magnetic impurity $\mathbf{M}_c$ can also point to arbitrary directions shown in the left inset, defined by the polar angle $\varphi$ and the azimuthal angle $\theta$.
    }
    \label{Fig1}
    \end{figure}

We consider the two-dimensional Josephson junction device in the $x$-$y$ plane with two magnetic impurities, as shown in Fig. \ref{Fig1}.
The length of the normal conductor area is $X$ and is limited in the region $-\frac{X}{2}<x<\frac{X}{2}$. The left semi-infinite superconductor area is in the region $x<-\frac{X}{2}$, and the right superconductor area is in $x>\frac{X}{2}$.
As for the magnetic impurities in the normal conductor area,
their direction and position can be adjusted.
In the continuum model, the Hamiltonian of this junction is written as:
\begin{eqnarray}
		H=\sum_{\mathbf{\hat{k}}} \Psi^{\dagger} \check{H}(\mathbf{\hat{k}}) \Psi,
\end{eqnarray}
with $\Psi=(\Psi_{\uparrow },\Psi_{\downarrow}, \Psi_{\uparrow}^{\dagger },\Psi_{\downarrow}^{\dagger } )^{T}$. $\Psi_{\alpha}^{\dagger }$ is the creation operator of electron with spin $\alpha$.
	
The $4 \times 4$ Bogoliubov-de Gennes (BdG) Hamiltonian is:
	\begin{eqnarray}
		\check{ H}(\mathbf{\hat{k}})=\begin{pmatrix}
			h(\mathbf{\hat{k}})   &  \Delta(x) i \sigma_y \\
			-\Delta^{*}(x) i \sigma_y & -h^*(-\mathbf{\hat{k}}) \\
		\end{pmatrix}, \label{E2}
\end{eqnarray}
with
\begin{eqnarray}
h(\mathbf{\hat{k}})=\frac{\hbar^2 \mathbf{\hat{k}}^2}{2m}+\bm{\sigma}\cdot \mathbf{M}-\mu(x). \label{E2a}
\end{eqnarray}
Here, $m$ is the effective mass and $\bm{\sigma} = (\sigma_{x}, \sigma_{y}, \sigma_{z})$ are the Pauli matrices.
$\mathbf{\hat{k}}=(\hat{k}_x, \hat{k}_y)$ is the two-dimensional derivation operator with $\hat{k}_x=-i\frac{\partial}{\partial x}$, $\hat{k}_y=-i\frac{\partial}{\partial y}$, and we take $\hbar=1$.

$\mathbf{M}$ in Eq. (\ref{E2a}) represents the exchange interaction between conduction electron and the magnetic impurity.
We emphasize that it is usually not caused by the magnetic field, which just flips the direction of the magnetic moment and is fundamentally different.
$\mathbf{M}$ is an intrinsic property originating from the electron-electron interaction in the magnetic impurity, which results in the spontaneous magnetization of impurity and is effectively transformed as the magnetic exchange \cite{phillips_advanced_2012}.
The magnitude of $\mathbf{M}$ is dependent on the charging energy and its coupling strength to conduction electron, of which the wide variation \cite{barreiro_quantum_2012,shibata_single_2023,kano_control_2015} leads to a broad range of $\mathbf{M}$ magnitude.
Thus, it covers the parameter range in our manuscript and better demonstrates the feasibility of experiment.

Here we set $\mathbf{M}=\mathbf{M}_{c} \delta(x,y)+\mathbf{M}_{s} \delta(x-x_s,y-y_s)$. $\mathbf{M}_{c}$ represents the magnetic moment of the center magnetic impurity, which always locates at position $(0,0)$.
The direction of $\mathbf{M}_{c}$ can be expressed by polar angle $\theta$ and azimuthal angle $\varphi$ as $\mathbf{M}_{c}=M \cdot (\sin \theta \cos \varphi, \sin \theta \sin \varphi, \cos \theta)$, as shown in the inset of Fig. \ref{Fig1}.
$\mathbf{M}_{s}$ represents the magnetic moment of the side magnetic impurity, which has an adjustable location $(x_s,y_s)$.
The side magnetic impurity can be magnetically soft.
Soft magnetic materials have broad domain wall and small coercive field which thus can be easily flipped in the direction of the magnetic field instantly \cite{fiorillo_hysteresis_2006,blundell_magnetism_2003,santhoshkumar_recent_2022}. The magnetic moment of the side magnetic impurity is flipped by the current-induced magnetic field in our paper (the feasibility of the flipping will be explained later in Sec. \ref{sec3})
, which can be at $+z$ and $-z$ directions depended on the current direction.
Thus, there are only two directions for the side magnetic moment, $+z$ and $-z$ directions, similar to the Ising magnetic moment.
So, $\mathbf{M}_{s}$ can be expressed by a sign function of current $I$, $\mathbf{M}_{s}=|M| \cdot (0,0,sgn(I))$.
Here we have assumed that the magnitudes $M$ of the magnetic moments of the central magnetic impurity and the side magnetic impurity are the same
and are not affected by the current-induced magnetic field.
This is equivalent to considering two identical magnetic impurities.
The results remain qualitatively unchanged even with two different magnetic impurities.

Besides, the chemical potential in Eq.(\ref{E2a})
and the s-wave superconducting gap in Eq.(\ref{E2}) are denoted as:
\begin{eqnarray}
		\mu(x) & = & \left\{\begin{matrix}
			\mu_{L} & x<-\frac{X}{2} \\
			\mu_N & -\frac{X}{2}<x<\frac{X}{2} \\
			\mu_{R} & x>\frac{X}{2}
		\end{matrix}\right.
	,\end{eqnarray}
	\begin{eqnarray}
		\Delta(x) =\left\{\begin{matrix}
			\Delta & x<-\frac{X}{2} \\
			0 & -\frac{X}{2}<x<\frac{X}{2} \\
			\Delta e^{i\Phi }  & x>\frac{X}{2}
		\end{matrix}\right. \label{E4}
	,\end{eqnarray}
with $\Delta=1$, $\mu_{L}=\mu_{R}=50$ and $\mu_N=20$ \cite{bagwell_suppression_1992,cayao_enhancing_2024} in the numerical calculations. $\Phi$ is the superconducting phase difference.
	
\begin{figure}[!htb]
    \centerline{\includegraphics[width=\columnwidth]{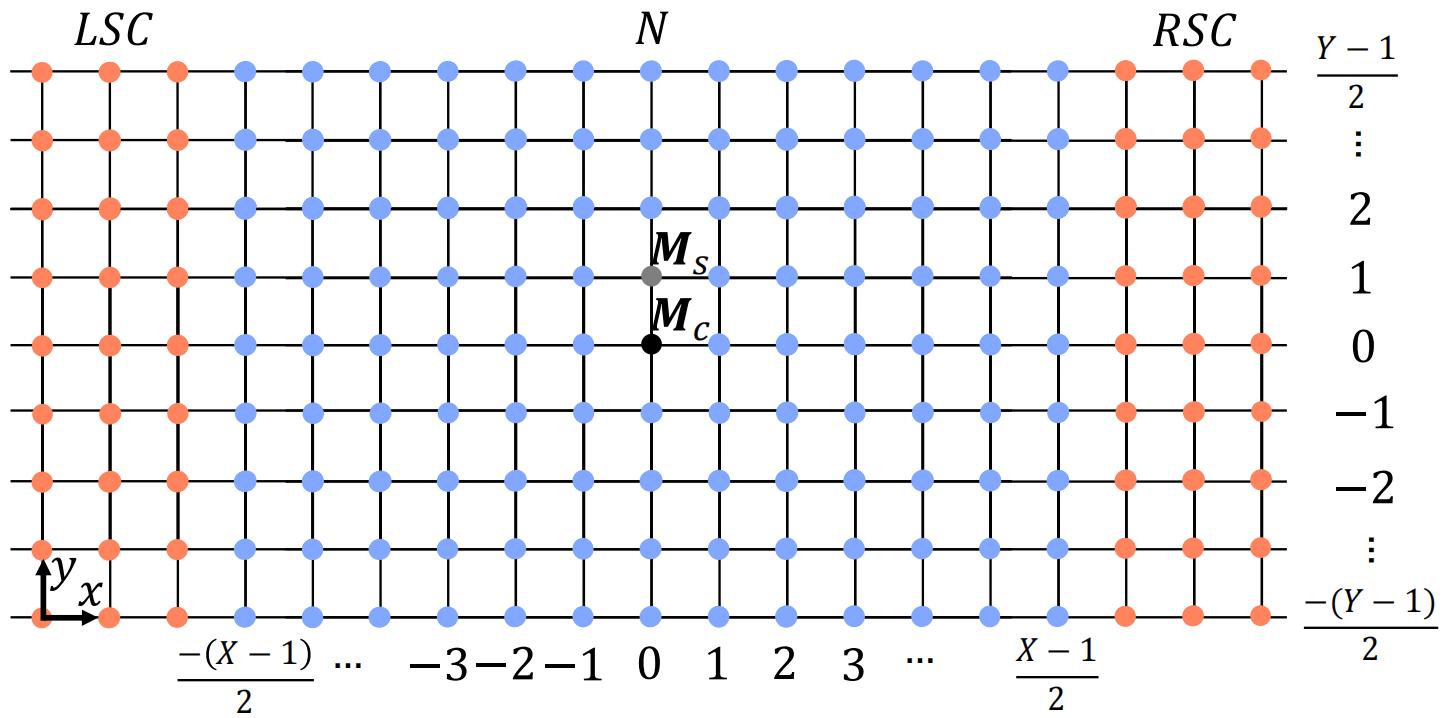}}
    \caption{
    Schematic illustration of the two-dimensional square lattice model obtained from the continuum model in Fig. \ref{Fig1}.
	The center magnetic moment locates at the coordinate origin $\left(x_c,y_c\right)=\left(0,0\right)$, which is fixed in the whole paper. The position of the side magnetic moment is $\left(x_s,y_s\right)$, which is $\left(0,1\right)$ in this figure. The lattice constant is $a$. the length of the normal conductor is $X$, and the width of the lattice is $Y$.
    }
    \label{Fig2}
    \end{figure}

To calculate the Josephson current of this model, we discretize this continuum Hamiltonian into a two-dimensional square lattice model as shown in Fig. \ref{Fig2}.
The site number along the $y$ direction of the junction is $Y$.
The site number of normal conductor along $x$ direction is $X$ and is limited in the region $-\frac{X-1}{2} \leqslant x \leqslant \frac{X-1}{2}$.
The left (right) semi-infinite superconductor area is placed in the region $x \leqslant -\frac{X+1}{2}$ ($x \geqslant \frac{X+1}{2}$).
The center magnetic impurity and the side magnetic impurity are placed at sites $(0, 0)$ and $(x_s, y_s)$, respectively.

The whole Josephson junction is discretized as \cite{sun_quantum_2005,cheng_spintriplet_2021,beenakker_universal_1991}:
	\begin{equation}
		H_{dis}=H_{LSC}+H_{RSC}+H_{N}+H_{T}+H_{M}
	.\end{equation}

The detailed discretization of Hamiltonians for the left (right) superconductor, the center normal conductor, and the tunneling Hamiltonian are shown in the Appendix.

	To calculate Josephson current, we define the particle number operator of electrons in left superconductor as \cite{sun_resonant_1999}:
	\begin{equation}
		N_L= \sum_{\substack{i_{x} \leqslant \frac{-(X+1)}{2}  \\
				\frac{-(Y-1)}{2}  \leqslant i_{y} \leqslant \frac{Y-1}{2}}}
		\sum_{\alpha } \Psi_{L i \alpha}^{ \dagger } \Psi_{L i \alpha}
	.\end{equation}
	The current can be calculated from the evolution of this operator as \cite{zhu_probing_2002,yan_anomalous_2020}:
	\begin{equation}
		\begin{aligned}
			I & = -e\left \langle \frac{\mathrm{d} N_L}{\mathrm{d} t}  \right \rangle = ie\left \langle \left [
			N_L, H_{dis} \right ]  \right \rangle  \\
			& = e \sum_{i_y} \mathrm{Tr} \left \{ \Gamma_z[\mathbf{G}_{NS}^<(t,i_y,t,i_y) \check{T}-\check{T}^{\dagger}\mathbf{G}_{SN}^<(t,i_y,t,i_y)] \right \} ,
		\end{aligned}  \label{E12}
	\end{equation}
	with $\Gamma_z=\mathrm{diag}(1,1,-1,-1)$.
	The lesser Green's function is defined as:
	\begin{equation}
		\mathbf{G}_{NS}^<(t,i_y,t^{\prime },i_y^{\prime })=i\left \langle \Psi_{\left (  \frac{-(X+1)}{2}, i_y^{\prime }\right ) }^{L \dagger }(t^{\prime })
		\otimes
		\Psi_{ \left ( \frac{-(X-1)}{2}, i_y \right ) }^{N} (t) \right \rangle .
        \end{equation}
The lesser Green's function can be Fourier transformed to the energy space $\mathbf{G}^{<}(t,t)=\int \frac{\mathrm{d} \varepsilon }{2\pi } \mathbf{G}^{<}(\varepsilon ) $, and $\mathbf{G}^{<}(\varepsilon )$ can be obtained by the fluctuation-dissipation theorem: $\mathbf{G}^{<}(\varepsilon )=-f(\varepsilon )[\mathbf{G}^{r}(\varepsilon )-\mathbf{G}^{a}(\varepsilon )]$ with the Fermi distribution $f(\varepsilon )$.
	$\mathbf{G}^{r}_{NS}$ ($\mathbf{G}^{a}_{NS}$) and $\mathbf{G}^{r}_{SN}$ ($\mathbf{G}^{a}_{SN}$) are the retarded (advanced) Green's functions. The retarded Green's functions can be obtained by the Dyson equation as:  $\mathbf{G}_{SN}^{r}=\mathbf{g} _{LS}^{r} \bm{\Sigma}_{SN}^{r} \mathbf{G}_{N}^{r}$ and $\mathbf{G}_{NS}^{r}=\mathbf{G} _{N}^{r}\bm{\Sigma}_{NS}^{r} \mathbf{g}_{LS}^{r}$ with $\bm{\Sigma}_{SN}^{r}=\check{T}$ and $\bm{\Sigma}_{NS}^{r}=\check{T}^{\dagger}$.
	Here, $\mathbf{g} _{LS}^{r}$ is the surface Green's function for the left superconductor which can be calculated by the transformation matrix method \cite{PhysRevB.93.195302,cheng_spintriplet_2021}. $\mathbf{G}_{N}^{r}$ is the Green's function of leftmost layer of the normal conductor region ($i_x= \frac{-(X-1)}{2}$), and it can be calculated by a recursive algorithm \cite{cheng_spintriplet_2021}. The advanced Green's functions are $\mathbf{G}^{a}_{NS}=[\mathbf{G}^{r}_{SN}]^{\dagger}$ and $\mathbf{G}^{a}_{SN}=[\mathbf{G}^{r}_{NS}]^{\dagger}$.
	
	By substituting these functions into Eq. (\ref{E12}), we can obtain the Josephson current as:
    \begin{equation}
        I=\frac{e}{2 \pi} \int d\varepsilon\ i\left(\varepsilon\right),  \label{E14}
    \end{equation}
    in which the $i(\varepsilon)$ is the integrand of the current in the negative-energy space \cite{mao_universal_2024}:
	\begin{equation}
		\begin{aligned}
			 i\left(\varepsilon\right)
			&=\mathrm{Tr}
			\{\Gamma_z [\mathbf{G}_{N}^{r}(\varepsilon) \bm{\Sigma}_{LS}^{<}(\varepsilon)+\mathbf{G}_{N}^{<}(\varepsilon) \bm{\Sigma}_{LS}^{a}(\varepsilon) \\
			&-\bm{\Sigma}_{LS}^{r}(\varepsilon) \mathbf{G}_{N}^{<}(\varepsilon)-\bm{\Sigma}_{LS}^{<}(\varepsilon) \mathbf{G}_{N}^{a}(\varepsilon)   ]\},
		\end{aligned}
	\end{equation}
    with the self-energies given by $\bm{\Sigma}_{LS}^{r}(\varepsilon)=\check{T}^{\dagger} \mathbf{g}_{LS}^{r}(\varepsilon) \check{T}$, $\bm{\Sigma}_{LS}^{a}(\varepsilon)=\check{T}^{\dagger} [\mathbf{g}_{LS}^{r}(\varepsilon)]^{\dagger} \check{T}$, and $\bm{\Sigma}_{LS}^{<}(\varepsilon)=-f(\varepsilon)[\bm{\Sigma}_{LS}^{r}(\varepsilon)-\bm{\Sigma}_{LS}^{a}(\varepsilon)]$.
	
    The current comes from two parts of integration in Eq. (\ref{E14}). For $-\Delta<\epsilon<0$ inside the superconducting gap, integrand $i\left(\varepsilon\right)$ is distributed around the discretized Andreev bound states with energy $\varepsilon_i$, called discrete current $I_{dis}$. For $\epsilon<-\Delta$, the integrand $i\left(\varepsilon\right)$ is continuously distributed in the energy space, called continuous current $I_{con}$.
	The magnitude of the discrete current is proportional to the derivative of the Andreev bound state energy $\varepsilon_i$ with respect to the superconducting phase difference: $I_{dis}\propto\sum_{i}{f\left(\varepsilon_i\right)\frac{\partial \varepsilon_i}{\partial\Phi}}$ \cite{bagwell_suppression_1992,krichevsky_spectrum_2000,zhang_josephson_2013}. The total current consists of discrete and continuous currents, with the discrete part accounting for the majority.

The units of energy and current are chosen as $\Delta=1$ and $\frac{e \Delta}{\pi \hbar}=1 $, respectively.
If $\Delta=1$ is a conventional $2$ meV, the current unit $\frac{e \Delta}{\pi \hbar}$ is about $154$ nA.
In the following calculation, we take the coupling strength $t=50$, and the size $X=Y=11$. These parameters correspond to about 60 nm $\times$ 60 nm experimental setups \cite{golod_single_2015,levajac_impact_2023}, according to the effective electron mass $m^\ast=0.014m_e$ with the electron mass $m_e$.
The system we study is the 2DEG, which widely appears in various systems, such as two-dimensional materials, metal surface, and so on. 
The magnitude of the effective mass chosen in our calculation is about $10^{-2}m_e$, and can be found in $\mathrm{InSb}$ \cite{zhang_quantized_2011} and InAs \cite{chrestin_critical_1994,lee_contribution_2019,yuan_experimental_2020}.
In principle, the 2DEG system with larger or smaller effective mass can also be used to design the JDE.

To numerically calculate the results in our paper, the imaginary part of the energy $\varepsilon$ is mostly taken as $\xi=0.05$. The only exception is the calculation about the distribution of $i(\varepsilon)$ in Figs. \ref{FIG3}(c, d), where we take $\xi=0.02$ to make the plot of Andreev bound state clearer.
This broadening parameter $\xi$ is a frequently-used numerical technique \cite{cheng_spintriplet_2021,cheng_josephson_2023}, where the Andreev bound state-induced discretized current is broadened in energy $\xi$, and it almost does not affect the calculated results.
What is more, the temperature which appears in the Fermi distribution is set to zero. The nonzero temperature has little influence on the result.

        \section{\label{sec3}EMERGENCE OF JDE}
	
As the magnetic effect of electric current, flowing currents can generate a magnetic field, which flips the magnetic moment of magnetic impurity.
For the magnetic impurity set at the center of the 2-dimensional device,
the currents flowing on its two sides are almost equivalent, and their generated magnetic fields are cancelled out.
	Thus, the magnetic moment of the center magnetic impurity feels no magnetic field and is not affected by currents.
	On the contrary, for the side magnetic impurity, the inequivalent currents on its two sides brings a finite net magnetic field, so its magnetic moment feels the magnetic field and gets flipped.

Note that the Josephson current $I$ is derived from the Hamiltonian,
meanwhile in the Hamiltonian [see Eq. (\ref{E2})] the side magnetic moment $\mathbf{M}_{s}=|M| (0,0,sgn(I))$ is a function related to the direction of current $I$.
For the positive current, the side magnetic moment is $\uparrow (+z)$-direction with $\mathbf{M}_{s}=|M| (0,0,1)$, while for the negative current, the side magnetic moment is $\downarrow (-z)$-direction with $\mathbf{M}_{s}=|M| (0,0,-1)$.
The direction of the side magnetic moment and the direction of the current influence each other,
and the calculated results are direction-dependent.
We can qualitatively notice that the positive and negative Josephson currents have different effects on the side magnetic moment and Hamiltonian, resulting in the difference of the values of the positive and negative current. Thus, the supercurrent is nonreciprocal and JDE appears.

In the quantitative calculation of this section,
we first set the direction of the center magnetic moment to point at $\left(\varphi,\theta\right)=\left(0,0\right)$, which is the $+z$ direction.
	The position of the side magnetic impurity is fixed at $\left(x_s,y_s\right)=\left(0,1\right)$ as shown in Fig. \ref{Fig2}.
	The magnitudes of the two magnetic moments are set as $M=80$.

    \begin{figure}[!htb]
    \centerline{\includegraphics[width=\columnwidth]{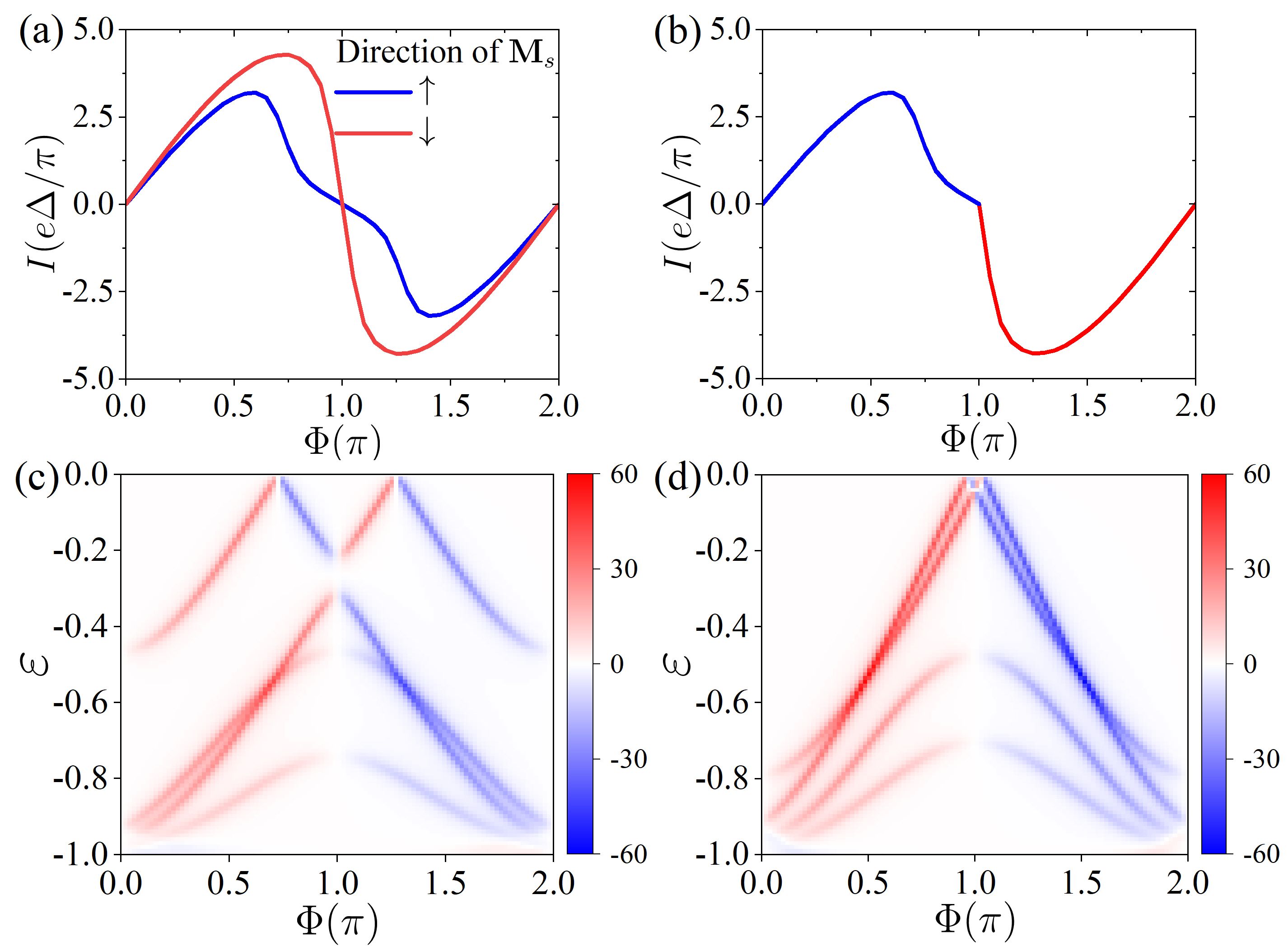}}
    \caption{
	(a) The CPRs for a fixed direction of side magnetic moment.
	The blue and red curves respectively correspond to side magnetic moment directions $\uparrow$ and $\downarrow$.
	(b) The CPR for the direction of side magnetic moment
depending on the current direction.
	For the positive (negative) current, the side magnetic moment points to $\uparrow$ ($\downarrow$), the CPR is the positive part of blue curve (negative part of red curve) in (a).
	(c, d) The current integrand $i(\varepsilon)$ versus the
energy $\varepsilon$ and the superconducting phase difference $\Phi$}
in the conditions of blue and red curves in (a), respectively.
    \label{FIG3}
    \end{figure}

For a theoretically concise clarification, we first just neglect the magnetic effect of the current on the side magnetic moment, i.e., the direction of the side magnetic moment is fixed and not dependent on the direction of current.
In Fig. \ref{FIG3}(a), we plot the CPRs of two conditions that the direction of the side magnetic moment points at $\uparrow$ and $\downarrow$ directions (shown as the blue and red curves respectively).
The two CPRs are each anti-symmetric with $I(\Phi)=-I(-\Phi)$. However, they have different shapes and critical values. The anti-symmetric relation is reflected by the joint inversion operator:
$\mathcal{S}=\mathcal{R}U_1\left(-\frac{\Phi}{2}\right)$, where $U_1\left(-\frac{\Phi}{2}\right)$ is the $U_1$ gauge transformation with phase $-\frac{\Phi}{2}$, and $\mathcal{R}$ is the mirror reflection operator about the $y$-$z$ plane.
For fixed magnetic moment $M$, we can obtain that the Hamiltonian satisfies the relation $\mathcal{S}H\left(\Phi,\ M\right)\mathcal{S}^\dagger=H\left(-\Phi,\ \ M\right)$, and the current is reversed by the $\mathcal{S}$ operator.
	Therefore, the CPR satisfies $I\left(\Phi,\ M\right)=-I\left(-\Phi,\ M\right)$.
	The maximum and the minimum values of the CPR are the positive and negative critical currents respectively, recognized as $I_{c+}$ and $I_{c-}$.
	The anti-symmetric CPR determines the result $I_{c+}=|I_{c-}|$, i.e., there is no JDE for fixed side magnetic moment.
	However, for the direction-dependent solution with $\mathbf{M}_s$ not fixed, the anti-symmetric relations break and the JDE can appear.

For the side magnetic impurity, our parameter $y_s=1$ corresponds to $y_s=6$ nm away from the center magnetic impurity, and width of the center region is $Y=60$ nm \cite{golod_single_2015,levajac_impact_2023}, we can estimate the current-induced magnetic field at the position of the side magnetic impurity by using Biot-Savart Law.
In Fig. \ref{FIG3}(a), the current can almost reaches 5 times of $\frac{e\Delta}{\hbar\pi}$, and the current-induced magnetic field can be about
${10}^{-6}$ T. 
This value is much larger than the coercive field of some soft magnets, including nanometer-size nanocrystallines. 
The coercive fields of them can be as low as $10^{-7}$ T \cite{krishnan_fundamentals_2016,blundell_magnetism_2003,santhoshkumar_recent_2022}, which achieves the requirement in our model.
As a result, the side magnetic impurity can be easily flipped by the Josephson current-induced magnetic field instantly.

Next, we consider the magnetic effect of the current on the side magnetic moment, and the direction of the side magnetic moment is flipped by the current.
We demonstrate the direction-dependent CPR in Fig. \ref{FIG3}(b).
	For the positive current, the direction of the side magnetic moment points at $\uparrow$ direction,
	thus the direction-dependent CPR is the positive part ($I>0$) of blue curve in Fig. \ref{FIG3}(a).
	For the negative current, the direction of the side magnetic moment is flipped to point at $\downarrow$ direction, thus the direction-dependent CPR is the negative part ($I<0$) of red curve in Fig. \ref{FIG3}(a).
	Because the direction-dependent CPR originates from the two CPRs in Fig. \ref{FIG3}(a) with different critical currents, $I_{c+}\neq|I_{c-}|$, and JDE emerges.

The origin of the nonreciprocity of the direction-dependent CPR
can be explained by comparing the two cases in Fig. \ref{FIG3}(a).
	As shown in Figs. \ref{FIG3}(c, d), we plot the distributions of the Andreev bound states by plotting the current integrand $i\left(\varepsilon\right)$ versus energy $\varepsilon$ and superconducting phase $\Phi$.
	The blue color represents a negative integrand, while the red color represents a positive integrand. The figures demonstrate that the discretized current is proportional to the derivative of Andreev bound states.
Fig. \ref{FIG3}(c) corresponds to that the direction of the side magnetic moment $\mathbf{M}_s$ is fixed at $\uparrow$ direction.
Around the phase $\Phi=0.7\pi$, one Andreev bound state corresponding to the positive current integrand $i\left(\varepsilon\right)$ crosses zero energy and appears in the positive-energy region.
	Due to the particle-hole symmetry, one Andreev bound state corresponding to the negative current integrand $i\left(\varepsilon\right)$ appears in the negative-energy space.
	Besides, the other three Andreev bound states with positive current integrands $i\left(\varepsilon\right)$ have negative energy in the whole $0<\Phi<\pi$ space.
At this point, part of the positive current is cancelled out, although the total Josephson current is positive in $0<\Phi<\pi$, and the current has a shift at $\Phi \approx 0.7\pi$, as shown by the blue curve in Fig. \ref{FIG3}(a).
Because the distribution of Andreev bound state is symmetric about $\Phi=\pi$, Josephson current satisfies the relation: $I_{c+}=|I_{c-}|$.
On the other hand, when the direction of the side magnetic moment $\mathbf{M}_s$ is fixed at $\downarrow$ direction, for $0<\Phi<\pi$ there are four Andreev bound states with positive current integrands $i\left(\varepsilon\right)$ in the negative-energy space in Fig. \ref{FIG3}(d).
At this point, there is no positive current cancelled out, and the total Josephson current is positive in $0<\Phi<\pi$ without shift.
The different distributions of Andreev bound states have different contributions to the total Josephson currents,
so their critical currents are totally unequal for the side magnetic moment being at $\uparrow$ and $\downarrow$ directions.
Because the direction-dependent CPR comes from the two cases in Fig. \ref{FIG3}(a) with different critical currents, the JDE appears. 
This direction-dependent CPR is one of the approaches to achieve the JDE, as there are many mechanisms of JDE, such as the momentum of Cooper pair \cite{davydova_universal_2022,pal_josephson_2022}.

The appearance of the nonreciprocal currents can also be explained in a phenomenological way by illustrating the scattering of electrons.
For the convenience of description, we only discuss the scattering of spin-up electrons, and the scattering of spin-down electrons is similar.
When there is no magnetic impurity in the junction, we utilize two wavefunctions to show how electrons pass through the normal conductor: $\psi_{c}$ demonstrates electrons passing through the position of the center magnetic impurity $\left(0,0\right)$, whereas $\psi_{s}$ demonstrates the electrons passing through the position of the side magnetic impurity $\left(x_s,y_s\right)$.
	
For a certain spin, the magnetic impurities can be regarded as delta potential barriers, with opposite signs for opposite directions of magnetic moments. When the directions of the center magnetic moment $\mathbf{M}_c$ and the side magnetic moment $\mathbf{M}_s$  are both pointed at $\uparrow$ direction, there are two delta potential barriers here as  $M\delta\left(x,y\right)$ and $M\delta\left(x-x_s,y-y_s\right)$.
	By the scattering of such delta potentials, the wavefunctions are reduced in amplitude by $\lambda$, and obtain an extra phase $\phi$. They can be written as:
    \begin{equation}
        \begin{aligned}                 \psi_{c}\rightarrow\lambda\psi_{c}e^{i\phi}, \
        \psi_{s}\rightarrow\lambda\psi_{s}e^{i\phi}.
        \end{aligned} \label{pathA}
    \end{equation}

When the direction of the center magnetic moment $\mathbf{M}_c$ is kept at $\uparrow$ direction, while the direction of the side magnetic moment $\mathbf{M}_s$ is flipped to $\downarrow$ direction, there is one delta potential barrier $M\delta\left(x,y\right)$ and one opposite delta potential barrier $-M\delta\left(x-x_s,y-y_s\right)$.
	For the two scatterings, the amplitude reduction is the same, while the phase differences of these two wavefunctions are opposite, which can be written as:
    \begin{equation}
        \begin{aligned}  \psi_{c}\rightarrow\lambda\psi_{c}e^{i\phi}, \
        \psi_{s}\rightarrow\lambda\psi_{s}e^{-i\phi}.
    \end{aligned} \label{pathB}
    \end{equation}

For electrons, the path through the center magnetic impurity and the path through the side magnetic impurity both contribute to the total wavefunction, and their interference influences the transmission probability of the electrons, which is the square modulus of total wavefunction. For the two magnetic moment configurations, the modulus of total wavefunctions are $|\lambda\left(\psi_{c}e^{i\phi}+\psi_{s}e^{i\phi}\right)|$ and $|\lambda\left(\psi_{c}e^{i\phi}+\psi_{s}e^{-i\phi}\right)|$, respectively, and they are unequal.
    Different electron transmission probability in the normal conductor leads to different tunneling transmission probability of Cooper pairs from the left superconductor to the right superconductor. Thus, the different magnetic configurations corresponding to opposite currents cause different critical currents. As a result, JDE emerges.

	\section{\label{sec4} THE DEPENDENCE OF THE JDE EFFICIENCY ON SYSTEM PARAMETERS }

To quantitatively study how JDE depends on a series of parameters in our junction, we introduce the superconducting diode efficiency $\eta$ to demonstrate the nonreciprocity of the critical currents:
\begin{equation}
\eta=\frac{{I}_{c+}-\left|I_{c-}\right|}{{I}_{c+}+\left|I_{c-}\right|}, \label{E18}
\end{equation}
For the direction-dependent Josephson CPR with $M=80$ shown in Fig. \ref{FIG3}(b), the superconducting diode efficiency $\eta$ can reach -14.5$\%$, which indicates the high nonreciprocity of our junction.

\begin{figure}[!htb]
    \centerline{\includegraphics[width=\columnwidth]{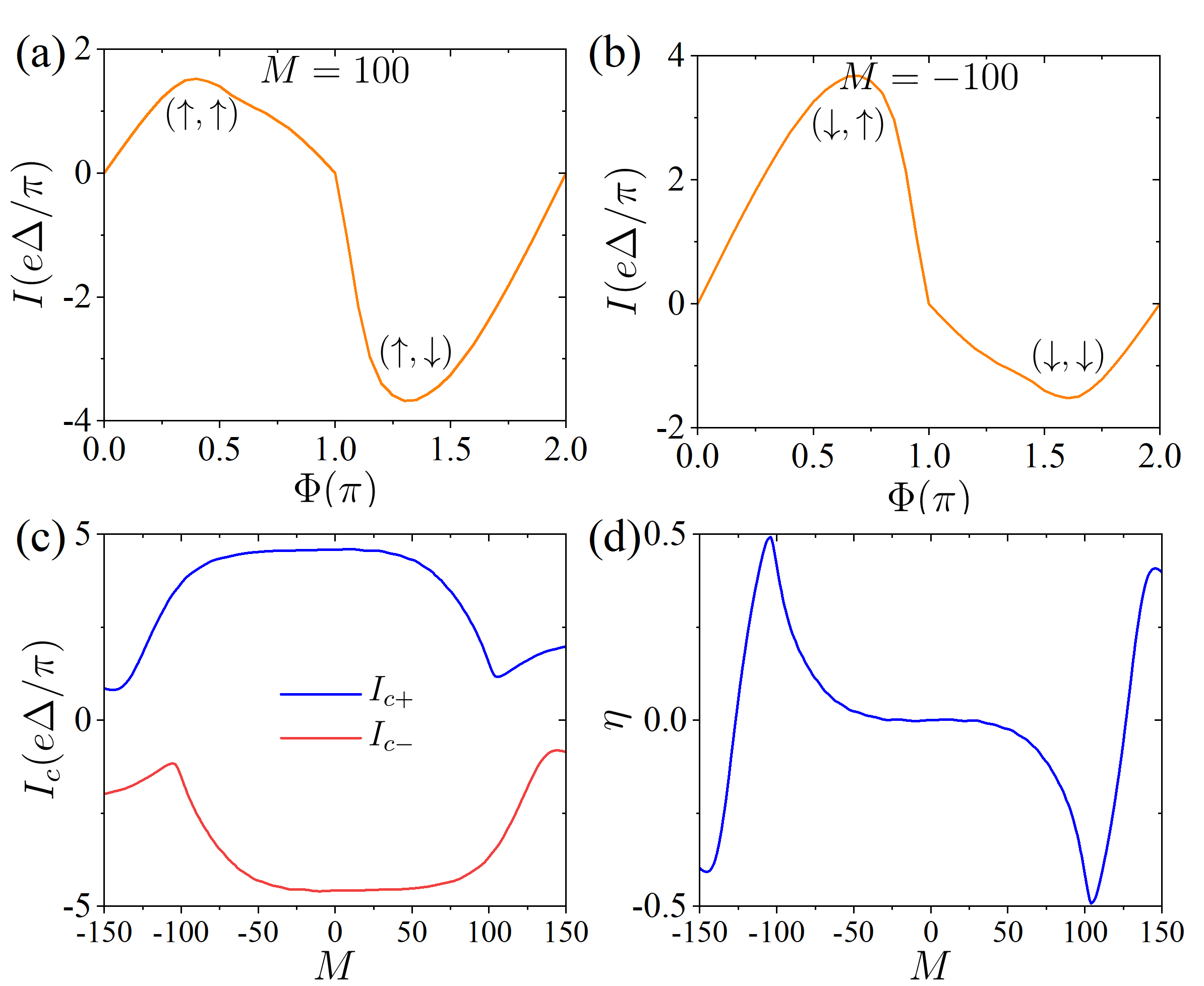}}
    \caption{(a, b) The direction-dependent CPRs for opposite directions of center magnetic moment. $\mathbf{M}_c$ points at $+z$ direction with $M=100$ in (a), while $\mathbf{M}_c$ points at $-z$ direction with $M=-100$ in (b).
	(c) Positive and negative critical currents $I_{c+}$ (blue curve) and $I_{c-}$(red curve) versus the magnitude of magnetic moment $M$.
	(d) Superconducting diode efficiency $\eta$ versus $M$.
	The positions of $\mathbf{M}_c$ and $\mathbf{M}_s$ are the same as those in Fig. \ref{Fig2}.
    }
    \label{FIG4}
    \end{figure}

First, we investigate the influence of the magnitude of the magnetic moments on the system.
The center magnetic moment is still fixed in direction but with a variable magnitude $M$, $\mathbf{M}_{c}=M \cdot (0, 0, 1)$.
Like the setting in Sec. \ref{sec2}, the side magnetic moment is determined by the current $\mathbf{M}_{s}=|M| \cdot (0,0,sgn(I))$.
For the two opposite directions of center magnetic moment $M=100, -100$, the direction-dependent CPRs are respectively shown in Figs. \ref{FIG4}(a, b).
The two CPRs both exhibit the JDE, but with opposite efficiency $\eta$.
For a given center magnetic moment direction $\mathbf{M}_{c} = \uparrow $ or $\downarrow$, the positive and negative currents correspond to $\mathbf{M}_{s} = \uparrow$ and $\downarrow$ respectively.
Thus, the direction-dependent CPRs originate from four types of magnetic configurations $(\mathbf{M}_{c}, \mathbf{M}_{s})=(\uparrow/\downarrow,\uparrow/\downarrow)$, as labelled in Figs. \ref{FIG4}(a, b).
It seems that the two CPR curves are centrosymmetric with respect to phase $\Phi=\pi$, which can be explained by the following symmetry relation.

    By using the time-reversal operator $\mathcal{T}=i\sigma_yK$ with $K$ being the complex conjugation operator, the Hamiltonian is transformed as $\mathcal{T}H\left(\Phi,\mathbf{M}_{c}, \mathbf{M}_{s}\right)\mathcal{T}^\dag=H\left(-\Phi,-\mathbf{M}_{c}, \mathbf{M}_{s}\right)$.
	Here $\mathbf{M}_{s}$ is first directly reversed by $\mathcal{T}$.
	Note that the current $I$ is also reversed, thus $\mathbf{M}_{s}$ is reversed again.
	Focusing on the two essential variables $\Phi$ and $M$, the transformation is equivalent to $\mathcal{T}H\left(\Phi,M\right)\mathcal{T}^\dag=H\left(-\Phi,-M\right)$.
	As a result, the CPR satisfies
	\begin{equation}I\left(\Phi,M\right)=-I\left(-\Phi,-M\right).\label{IM}\end{equation}

    In Figs. \ref{FIG4}(c, d), we demonstrate the positive and negative critical currents $I_{c+}$ and $I_{c-}$ versus the magnitude of magnetic moment $M$, as well as the efficiency $\eta$.
	The nonreciprocity is absent for $M=0$, but remarkably emerges as the magnetic moment is included.
	The efficiency can achieve $\eta\approx50\%$ near $M=100$.
	According to the relation Eq. (\ref{IM}),
	we can easily find $I_{c+}\left(M\right)=\left|I_{c-}\left(-M\right)\right|$.
	Also, the superconducting diode efficiency $\eta$ is an odd function of the magnetic moment $M$, with $\eta\left(M\right)=-\eta\left(-M\right)$.

\begin{figure}[!htb]
    \centerline{\includegraphics[width=\columnwidth]{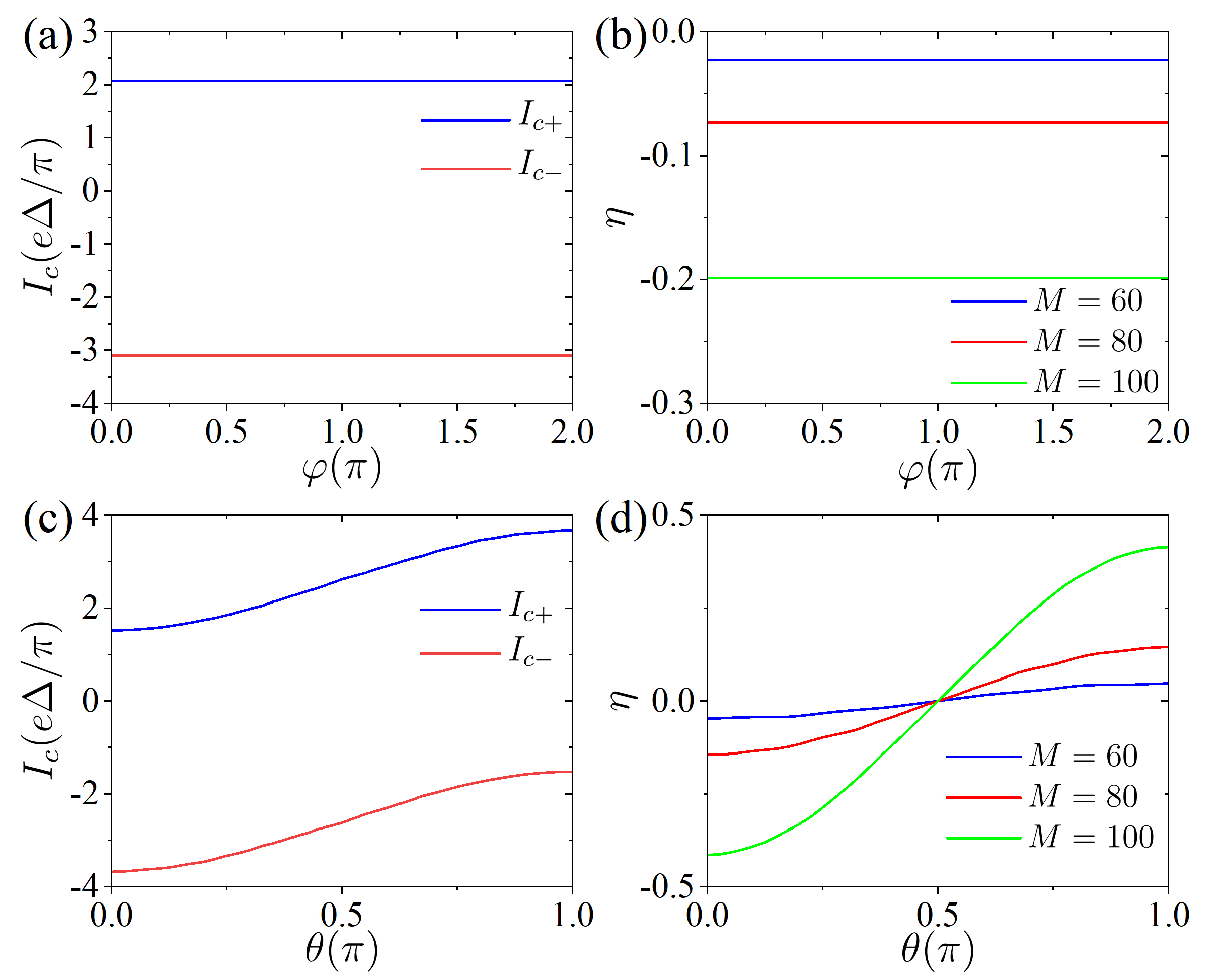}}
    \caption{(a, c) Positive and negative critical currents $I_{c+}$ (blue curve) and $I_{c-}$ (red curve) versus the polar angle $\varphi$ and the azimuthal angle $\theta$, respectively.
    The magnitude of the magnetic moments is $M=100$.
    (b, d) Superconducting diode efficiency $\eta$ versus the polar angle $\varphi$ and the azimuthal angle $\theta$ for various magnitudes of magnetic moment $M$.
	Besides, the azimuthal angle is fixed at $\theta=\pi/3$ in (a, b), while the polar angle is fixed at $\varphi=0$ in (c, d). The positions of $\mathbf{M}_c$ and $\mathbf{M}_s$ are the same as those in Fig. \ref{Fig2}.
    }
    \label{FIG5}
    \end{figure}

Next, we analyze the influence of the direction of the center magnetic moment on the system, which is described by the azimuthal angle $\theta$ and the polar angle $\varphi$.
In Fig. \ref{FIG5}(a), we plot the positive and negative critical currents versus the polar angle $\varphi$ of the center magnetic moment, with the fixed azimuthal angle $\theta=\pi/3$.
The critical currents are two unequal constants independent on $\varphi$.
To analyze the symmetry about the directions of magnetic moments, we express the Hamiltonian as $H(\varphi,\theta,\mathbf{M}_{s})$.
	By using the spin rotation operator $U_{s_z}=e^{i\frac{\varphi^\prime}{2}\sigma_z}$ by angle $\varphi^\prime$ about the $z$ axis, the Hamiltonian is transformed as: $U_{s_z} H(\varphi,\theta,\mathbf{M}_{s}) U_{s_z}^\dag=H(\varphi-\varphi^\prime,\theta,\mathbf{M}_{s})$, while the current $I$ is unchanged.
	Thus, the current satisfies the relation $I\left(\varphi\right)=I\left(\varphi-\varphi^\prime\right)$, indeed independent on the polar angle of center magnetic moment $\varphi$.
	As a result, the efficiency $\eta$ is also invariant with the change in polar angle, as shown by curves with different $M$ values in Fig. \ref{FIG5}(b).

	We also plot the positive and negative critical currents versus the azimuthal angle $\theta$ of the center magnetic moment with the polar angle $\varphi=0$, as shown in Fig. \ref{FIG5}(c).
	As $\theta$ increases from $0$ to $\pi$, $I_{c+}$ increases but $|I_{c-}|$ decreases.
	Using the spin rotation operator $U_{s_x}=e^{i\frac{\pi}{2}\sigma_x}$, the Hamiltonian is transformed as $U_{s_x}H\left(\varphi,\theta,\mathbf{M}_{s}\right)U_{s_x}^\dag=H\left(\varphi,\pi-\theta,-\mathbf{M}_{s}\right)$.
	Note that the reversion of $\mathbf{M}_{s}$ corresponds to the reversion of current $I$, thus
	the critical currents satisfy the relation $I_{c+}\left(\theta\right)=-I_{c-}\left(\pi-\theta\right)$.
	Consequently, the efficiency $\eta$ satisfies $\eta\left(\theta\right)=-\eta\left(\pi-\theta\right)$, anti-symmetric with respect to the angle $\theta=\pi/2$ as shown in Fig. \ref{FIG5}(d).
	The results show that the JDE can significantly appear in a wide range of parameters.
	Even if the direction of center magnetic moment is a bit deviated from $z$ direction, the efficiency $\eta$ can keep a relatively high value.
	
	\begin{figure}
		\centerline{\includegraphics[width=\columnwidth]{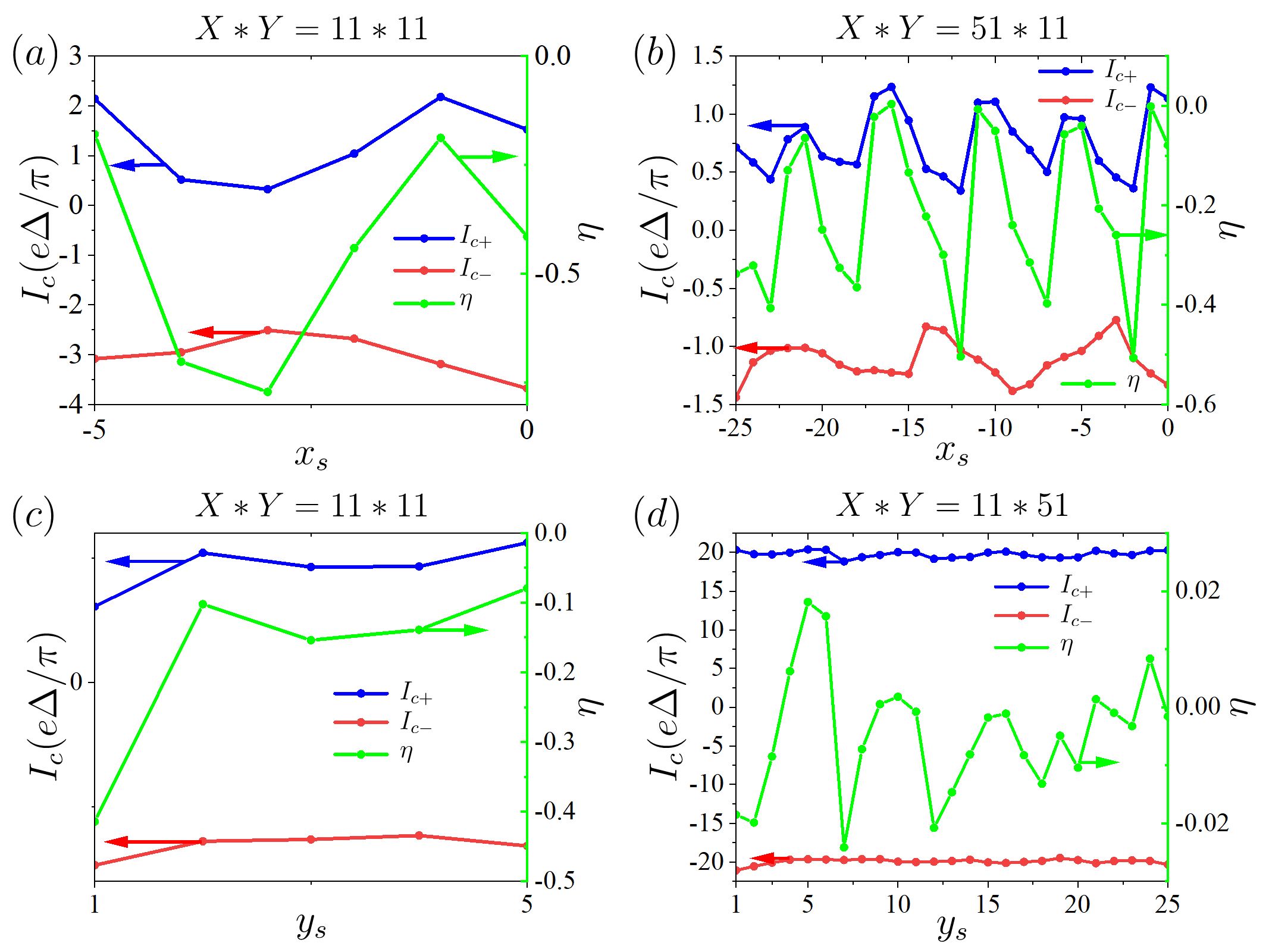}}
		\caption{Positive critical current $I_{c+}$ (blue curve), negative critical current $I_{c-}$ (red curve), and superconducting diode efficiency $\eta$ (green curve) versus the position of the side magnetic moment $\left(x_s,y_s\right)$ in different size of systems. The system sizes are labeled above each subfigure. (a, b): $\left(x_s,y_s\right)$ moves horizontally from position (-5, 1) to (0, 1), and from position (-25, 1) to (0, 1), respectively; (c, d): $\left(x_s,y_s\right)$ moves vertically from position (0, 1) to (0, 5), and from position (0, 1) to (0, 25), respectively. The magnitude of the magnetic moments is $M=100$. Besides, the direction of the center magnetic moment is fixed at $\left(\varphi,\theta\right)=\left(0,0\right)$.
	}
		\label{FIG6}
		\end{figure}

At last, we study the influence of the position of the side magnetic moment on the system, which is described by $\left(x_s,y_s\right)$ in Fig. \ref{Fig2}. 
Fig. \ref{FIG6}(a, c) shows the influence of the position of the side magnetic moment on our default system ($X \ast Y =11 \ast 11$), while Fig. \ref{FIG6}(b, d) shows the influence on larger systems.
As $y_s$ is fixed, we plot the positive critical current, negative critical currents, and the superconducting diode efficiency $\eta$ versus $x_s$ in Fig. \ref{FIG6}(a, b).
As $x_s$ is fixed, we plot the positive critical current, negative critical currents, and the superconducting diode efficiency $\eta$ versus $y_s$ in Fig. \ref{FIG6}(c, d).
It can be seen that the JDE is remarkable for a wide range of positions. Also, the critical current and superconducting diode efficiency curves oscillate with the position of the side magnetic moment $\left(x_s,y_s\right)$.
In Fig. \ref{FIG6}(a, b), the absolute efficiency is not maximized at $(x_s,y_s)=(0,1)$, the position nearest to the center magnetic moment.
	When the side magnetic moment is located at other places like $(-3, 1)$
	in Fig. \ref{FIG6}(a) and $(-2, 1)$ in Fig. \ref{FIG6}(b),
	the absolute efficiency could even be much higher.
	In fact, the oscillation originates from the wavefunction modulation of electrons passing through the connecting region, as shown in Eqs. (\ref{pathA}, \ref{pathB}).
	The nonreciprocity comes from the different interference of two magnetic configurations
	$|\lambda\left(\psi_{c}e^{i\phi}+\psi_{s}e^{i\phi}\right)|$ and $|\lambda\left(\psi_{c}e^{i\phi}+\psi_{s}e^{-i\phi}\right)|$.
	Here $\psi_c$ ($\psi_s$) is the wavefunction passing from one superconductor through the center (side) magnetic moment position to the other superconductor.
	As the position of side magnetic moment is changed, the phase accumulation of $\psi_s$ is changed due to the change of path length.
	As a result, the interference between the two paths (coherence or cancellation) is modulated by the position of the side magnetic moment, causing the curves to oscillate with the change of $\left(x_s,y_s\right)$.
The overarching trend of interference can be seen in Fig. \ref{FIG6}(b, d), the calculations for larger systems. The corresponding critical currents and efficiency versus impurity position exhibit obvious oscillation patterns. They better support the statement about interference.

    \section{\label{sec5}DISCUSSION AND CONCLUSION}

In conclusion, we propose a theoretically concise and universal superconducting diode device based on magnetic atoms.
In the presence of positive and negative currents, opposite magnetic fields are generated, and the magnetic configurations are different, in turn leading to unequal positive and negative critical currents, which is called the JDE.
We calculate the CPRs and confirm the existence of JDE by the nonequilibrium Green's function.
We quantitatively and phenomenologically explain the mechanism of JDE.
Furthermore, we investigate the superconducting diode efficiency in a wide parameter space, including the magnitude $M$, the direction $\left(\theta,\varphi\right)$, and the position $\left(x_s,y_s\right)$ of the magnetic moment.
We also analyze the symmetry relations between them in detail.
Our work has high efficiency and flexible controllability, with no demand on
the external magnetic field, bulk Cooper pair momentum, or the spin-orbit coupling.
Our proposal gives a new perspective on the development of superconducting devices and paves the way for the construction of nonreciprocal electronic circuits.

  \section*{\label{sec6}ACKNOWLEDGMENTS}
This work was financially supported
        by NSF-China (Grants No. 11921005 and No. 12374034),
        the Innovation Program for Quantum Science and
        Technology (No. 2021ZD0302403), and 
		the National Key R and D Program of China (Grant No. 024YFA1409002). 
		The computational
        resources are supported by the High-Performance
        Computing Platform of Peking University.	

\appendix
\section*{Appendix: Discretization of the Hamiltonian}

\begin{widetext}	
 The discrete Hamiltonians for left and right superconductors $H_{LSC}$ and $H_{RSC}$:
	\begin{eqnarray}
			H_{LSC} &=& \sum_{\substack{i_{x} \leqslant\frac{-(X+1)}{2}  \\
					\frac{-(Y-1)}{2}  \leqslant i_{y} \leqslant \frac{Y-1}{2}}}
			\Psi_{i}^{L \dagger } \check{H}_{0}^{L} \Psi_{i}^{L}
			 +
\left[\sum_{\substack{i_{x} \leqslant \frac{-(X+3)}{2} \\
					\frac{-(Y-1)}{2}  \leqslant i_{y} \leqslant \frac{Y-1}{2}}}
			\Psi_{i}^{L \dagger } \check{H}_{x}^{L} \Psi_{i+\delta x}^{L}
			 +\sum_{\substack{i_{x} \leqslant \frac{-(X+1)}{2} \\
					\frac{-(Y-1)}{2}  \leqslant i_{y} \leqslant \frac{Y-3}{2}}}
			\Psi_{i}^{L \dagger } \check{H}_{y}^{L} \Psi_{i+\delta y}^{L}+\text { H.c. }\right],\\
			H_{RSC} &= &\sum_{\substack{i_{x} \geqslant  \frac{X+1}{2}  \\
					\frac{-(Y-1)}{2}  \leqslant i_{y} \leqslant \frac{Y-1}{2}}}
			\Psi_{i}^{R \dagger } \check{H}_{0}^{R} \Psi_{i}^{R}
			 +\left[\sum_{\substack{i_{x} \geqslant  \frac{X+1}{2} \\
					\frac{-(Y-1)}{2}  \leqslant i_{y} \leqslant \frac{Y-1}{2}}}
			\Psi_{i}^{R \dagger } \check{H}_{x}^{R} \Psi_{i+\delta x}^{R}
			 +\sum_{\substack{i_{x} \geqslant  \frac{X+1}{2}\\
					\frac{-(Y-1)}{2}  \leqslant i_{y} \leqslant \frac{Y-3}{2}}}
			\Psi_{i}^{R \dagger } \check{H}_{y}^{R} \Psi_{i+\delta y}^{R}+\text { H.c. }\right].
	\end{eqnarray}

Here, $\Psi_{i}^{L(R)}=(\Psi_{L(R)i \uparrow}, \Psi_{L(R)i \downarrow}, \Psi_{L(R)i \uparrow}^{\dagger },\Psi_{L(R)i \downarrow}^{\dagger } )^{T}$ in which $\Psi_{L(R)i \alpha}^{\dagger}$ is the creation operator of electron with spin $\alpha$ on lattice site $i=(i_x, i_y)$ in the left (right) superconductor.
	The Hamiltonian matrices are:
	\begin{equation}
		\check{H}_{0}^{L(R)}=\begin{pmatrix}
			h_{L(R)} & 0 & 0 & \Delta_{L(R)} \\
			0 & h_{L(R)} & -\Delta_{L(R)} & 0\\
			0 & -\Delta_{L(R)}^* & -h_{L(R)}^* & 0 \\
			\Delta_{L(R)}^* & 0 & 0 & -h_{L(R)}^*
		\end{pmatrix}
	,\end{equation}
with $h_{L(R)}=\frac{2}{ma^2}-\mu_{L(R)}$. Here $\Delta_{L}=\Delta$ and  $\Delta_{R}=\Delta e^{i\Phi }$, as shown in Eq. (\ref{E4}).
$\check{H}_{x}^{L(R)}= \check{H}_{y}^{L(R)}=\mathrm{diag} (-\frac{1}{2ma^2}, -\frac{1}{2ma^2}, \frac{1}{2ma^2}, \frac{1}{2ma^2})$
with $a$ the lattice constant.
	
The discrete Hamiltonian for center normal conductor is:
	\begin{equation}
			H_{N}  = \sum_{ \substack{ \frac{-(X-1)}{2} \leqslant i_{x} \leqslant   \frac{X-1}{2}  \\
					\frac{-(Y-1)}{2}  \leqslant i_{y} \leqslant \frac{Y-1}{2} }}
			\Psi_{i}^{N \dagger } \check{H}_{0}^{N} \Psi_{i}^{N}
			 +\left[\sum_{\substack{\frac{-(X-1)}{2} \leqslant i_{x} \leqslant   \frac{X-3}{2} \\
					\frac{-(Y-1)}{2}  \leqslant i_{y} \leqslant \frac{Y-1}{2}}}
			\Psi_{i}^{N \dagger } \check{H}_{x}^{N} \Psi_{i+\delta x}^{N}
+\sum_{\substack{\frac{-(X-1)}{2} \leqslant i_{x} \leqslant   \frac{X-1}{2} \\
					\frac{-(Y-1)}{2}  \leqslant i_{y} \leqslant \frac{Y-3}{2} }}
			\Psi_{i}^{N \dagger } \check{H}_{y}^{N} \Psi_{i+\delta y}^{N}+\text { H.c. }\right].
	\end{equation}
\end{widetext}

 Here, $\Psi_{i}^{N}=(\Psi_{N i \uparrow}, \Psi_{N i \downarrow}, \Psi_{N i \uparrow}^{\dagger },\Psi_{N i \downarrow}^{\dagger } )^{T}$ in which $\Psi_{N i \alpha}^{\dagger}$ is the creation operator of electron with spin $\alpha$ on lattice site $i=(i_x, i_y)$ in the normal conductor.
	The Hamiltonian matrices are:
	$\check{H}_{0}^{N}=\mathrm{diag} (\frac{2}{ma^2}-\mu_N, \frac{2}{ma^2}-\mu_N, -\frac{2}{ma^2}+\mu_N, -\frac{2}{ma^2}+\mu_N)$, and $\check{H}_{x}^{N}=\check{H}_{y}^{N}=\check{H}_{x}^{L(R)}$. 
	
The tunneling Hamiltonian between different regions are written as:
	\begin{equation}
		\begin{aligned}
			H_{T} & = \sum_{ \substack{\frac{-(Y-1)}{2}  \leqslant i_{y} \leqslant \frac{Y-1}{2} }}
			[\Psi_{\frac{-(X+1)}{2}}^{L \dagger } \check{T} \Psi_{\frac{-(X-1)}{2}}^{N} \\
			& + \Psi_{\frac{X+1}{2}}^{R \dagger } \check{T} \Psi_{\frac{X-1}{2}}^{N} +\text { H.c. }],
		\end{aligned}
	\end{equation}
with the hopping matrix $\check{T}=\text{diag} (t, t, -t^{*}, -t^{*})$. $t=\frac{1}{2ma^2}$ is the coupling strength between normal conductor and superconductors. For simplicity, the subscript $\frac{(X-1)}{2}$ of $\Psi_{\frac{(X-1)}{2}}^{N}$ denotes the site $i=(\frac{(X-1)}{2}, i_y)$, and other subscripts here have the same meaning.

    The Hamiltonian $H_{M}$ describes the two magnetic impurities
	\begin{equation}
			H_{M}  = \Psi_{i_c}^{N \dagger } \check{H}_{0}(\mathbf{M}_c) \Psi_{i_c}^{N}
			 + \Psi_{i_s}^{N \dagger } \check{H}_{0}(\mathbf{M}_s) \Psi_{i_s}^{N}.
	\end{equation}

 Here $i_c=(0,0)$ and $i_s=(x_s,y_s)$ are the positions of magnetic impurities.
	$\check{H}_{0}(\mathbf{m}) = \begin{pmatrix}
		\bm{\sigma}\cdot \mathbf{m}   &  0_{2 \times 2} \\
		0_{2 \times 2}  & -(\bm{\sigma}\cdot \mathbf{m})^* \\
	\end{pmatrix} $, with $\mathbf{m}=\mathbf{M}_c$ or $\mathbf{M}_s$.

\bibliography{reference.bib}

\end{document}